# Retrouver l'inventeur-auteur : la levée d'homonymies d'autorat entre les brevets et les publications scientifiques

David Reymond[1], Manuel Durand-Barthez[2], Heman Khouilla[3], Sandrine Wolff[4]

[1]Université de Toulon, Aix Marseille Univ, IMSIC - dreymond@univ-tln.fr

[2]URFIST Paris - manuel.durand-barthez@orange.fr

[3]Université de Toulon, LÉAD – heman-khouilla-koumba@etud.univ-tln.fr

[4]Université de Strasbourg, Université de Lorraine, CNRS, BETA – wolff@unistra.fr

## Abstract

Patents and scientific papers provide an essential source for measuring science and technology output, to be used as a basis for the most varied scientometric analyzes. Authors' and inventors' names are the key identifiers to carry out these analyses, which however, run up against the issue of disambiguation. By extension identifying inventors who are also academic authors is a non-trivial challenge. We propose a method using the International Patent Classification (IPC) and the IPCCAT API to assess the degree of similarity of patents and papers abstracts of a given inventor, in order to match both types of documents. The method is developed and manually qualified based on three corpora of patents extracted from the international EPO database Espacenet. Among a set of 4679 patents and 7720 inventors, we obtain 2501 authors. The proposed algorithm solves the general problem of disambiguation with an error rate lower than 5%.

## Résumé

Brevets d'invention et articles scientifiques représentent une source essentielle pour mesurer la production scientifique et technologique, servant de base aux analyses scientométriques les plus variées. Les noms d'auteurs et d'inventeurs constituent un identifiant pivot pour réaliser ces analyses, mais ils nécessitent d'être désambiguïsés. Par extension l'identification des inventeurs qui sont également des auteurs académiques est un problème non trivial. Nous proposons une méthode utilisant la classification internationale des brevets et l'API IPCCAT pour évaluer le degré de similarité entre le résumé du brevet d'un inventeur donné et ceux des articles de personnes homonymes, afin d'apparier des deux types de documents. La méthode est développée et qualifiée manuellement sur trois corpus de brevets issus de la base internationale Espacenet de l'OEB. Sur un jeu de 4679 brevets et 7720 inventeurs, nous obtenons un résultat de 2501 auteurs. L'algorithme proposé résout le problème général de la levée d'homonymie avec un taux d'erreur inférieur à 5%.

**Mots clés :** désambiguïsation ; auteurs ; inventeur ; homonymie ; brevets ; publications

## 1. Introduction

Les brevets d'invention comme les articles scientifiques, en tant que produits identifiables et mesurables des activités de recherche et développement, constituent la source informationnelle la plus répandue pour analyser la recherche et l'innovation, en particulier les interactions entre la Science et l'Industrie[1]. Les noms d'auteurs ou d'inventeurs constituent un identifiant pivot pour la réalisation d'analyses sur ces questions, mais ils comportent des problèmes de désambiguïsation notamment par rapport à l'homonymie, c'est-à-dire une même forme lexicale pour deux personnes différentes[2], étant entendu que le point de départ est constitué de bases de données accessibles librement et, en général, non nettoyées (Laender et al., 2008; Kim et al., 2014; Hussain & Asghar, 2017).

En utilisant la classification internationale des brevets (CIB[3]) et l'API IPCCAT comme instruments de mesure de la similarité des productions scientifiques et technologiques associées à un nom, nous proposons une nouvelle méthode de levée d'homonymie d'autorat, que nous expérimentons sur trois corpus de brevets extraits de la base internationale Espacenet de l'Office Européen des Brevets (OEB). La qualification des résultats de l'algorithme a été effectuée manuellement via la méthode d'échantillonnage statistique. Sur un ensemble de 4679 brevets correspondant à 7720 inventeurs, 2501 parmi eux sont aussi des auteurs. L'algorithme proposé résout le problème général de la levée d'homonymie avec un taux de réussite très satisfaisant (<5% d'erreur).

Au sein des trois corpus, notre objectif est d'identifier, parmi les noms d'inventeurs, ceux qui sont aussi des auteurs de publications, fort probablement issus du monde académique. La méthodologie est construite en trois phases. La première regroupe au niveau lexical les inventeurs. La seconde collecte dans une base complémentaire les données des publications académiques associées à chaque nom d'inventeur. Enfin, la dernière étape consiste à rapprocher le résumé des publications associées à ce nom (qu'il s'agisse d'un homonyme ou d'un potentiel inventeur-auteur) avec le descriptif du contenu du brevet, ce qui doit permettre *in fine* de déterminer si l'inventeur est bien l'auteur des publications.

## 2. Construction des jeux de données

Pour parvenir à notre objectif, nous avons construit à l'aide du logiciel *Patent2Net* (Reymond 2021) trois corpus de demandes de dépôts de brevets, issus de la base mondiale des brevets worldwide.espacenet.com hébergée par l'OEB. Les requêtes se limitent aux demandes de priorité française (PR=FR) et ciblent un domaine technologique particulier et une période[4].

- **Univ** : Vise des brevets détenus par des organismes de recherche publique ("univ* ou laboratoire ou institut" dans le nom du déposant) sur la période 2014-2016. La requête est : "(pa=univ* or pa=institut or pa=laboratoire) AND (ic=A61) AND (pd within "2014, 2016") AND pr=FR". Le domaine technologique choisi (A61) cible, dans la grande catégorie des nécessités courantes de la vie, le domaine médical ou vétérinaire et l'hygiène. On y verra majoritairement des déposants universitaires, acteurs de la valorisation du domaine.
- **Large** : Même période, mais ne cible pas des déposants spécifiques et vise deux domaines technologiques **différents** (A63 - sports, jeux…- et A62 – sauvetage ; lutte contre les incendies), sans imposer de contrainte sur le nom de déposant *a priori*. La requête est : "*(ic=A63 OR IC=A62) AND (pd within "2014, 2016") AND pr=FR*".
- **LargeBis** : Même domaine que **Univ**, mais sur une **autre période** proche (2013). La requête est : "ic=A61 AND pd=2013 AND pr=FR". Se distingue de **Univ** par la levée de la contrainte sur le nom de déposant.

---

[1] Cf. Foray & Lissoni (2010) pour une synthèse générale sur les relations universités-entreprises.
[2] La désambiguïsation des noms de déposants (Dassa *et al.* 2014) n'est pas abordée ici, car pour nos expériences elle est réalisée à l'aide d'un dictionnaire fourni par l'OEB, et une table de correspondance, que nous avons développée en extension de la ressource fournie par l'OEB (Office 2019).
[3] Ou *International Patent Classification* (IPC) en anglais.
[4] Les critères de sélection des corpus visent à lever les biais technologiques et temporels tout en appréciant l'efficacité de la méthode.

## 3. Procédure d'extraction

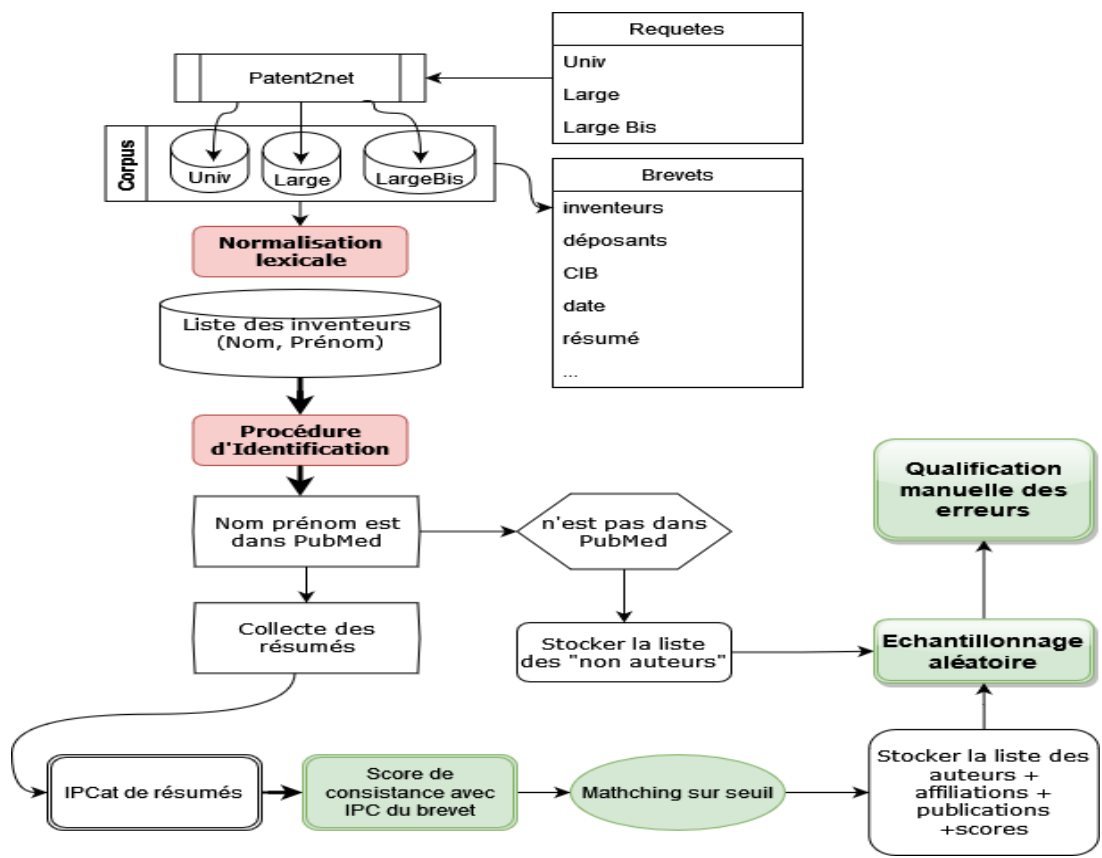

*Figure 1: Logigramme des différentes opérations réalisées, depuis les requêtes à l'API EspaceNet pour générer les trois corpus, la normalisation lexicale et la procédure d'identification, jusqu'à la qualification de la procédure de désambiguïsation des auteurs- inventeurs (via un échantillonnage aléatoire sur le corpus).*

La Figure 1 présente le fil des traitements réalisés sur les différents corpus est détaillé ci-après :

1. Une pré-normalisation s'appuie sur la bibliothèque python *fuzzy-wuzzy*[5] avec une distance de Levenshtein insensible à la position des mots pour rapprocher des chaînes lexicales (inversions noms-prénoms, coquilles). La chaîne la plus longue est privilégiée. Cette étape construit la liste des inventeurs.
2. Procédure d'identification de chaque inventeur en deux temps :
    a. Interrogation de l'API PubMed pour récupérer l'ensemble des publications associées au dit inventeur. L'outil de désambiguïsation de PubMed (Torvik *et al.* 2005 ; Lerchenmueller et Sorenson 2016) permet de chercher les variations d'une forme lexicale.

---

[5] Cf. https://github.com/seatgeek/thefuzz

      b. Chaque production identifiée est rattachée à l'inventeur. Cette étape produit la liste des auteurs homonymes aux inventeurs.
3. L'API IPCCAT (Fiévet et Guyot 2018a, 2018b) est ensuite utilisée ensuite de manière originale[6] pour rapprocher le résumé de chaque publication scientifique d'auteur homonyme au résumé du brevet correspondant, en indiquant pour chaque texte les classes technologiques de la CIB identifiées. Les scores de pertinence des CIB retournés par l'API sont alors utilisés pour associer les auteurs :
      a. Si le score est >800[7], et si l'un des codes CIB (5 digits) attribués au brevet de l'inventeur figure parmi ceux associés par IPCCAT aux publications scientifiques, alors l'inventeur est considéré comme l'auteur des publications.
      b. Dans ce cas, un appel à l'API de géoréférencement avec l'adresse fournie par Pubmed (pas toujours complète) vient compléter l'information auteur.
4. Une opération manuelle de vérification sur le Web (LinkedIn, Research Gate, Google Scholar…) est effectuée par la suite sur 10% des auteurs prélevés aléatoirement, afin de contrôler l'autorat.

Implanté en open source (https://github.com/Patent2net/P2N-v3/tree/Prg7), le procédé est flexible : la procédure d'identification (étape 2, figure 1 peut-être notamment adaptée à la collecte de publications issues de bases bibliographiques pluridisciplinaires (le Web of Science, Scopus, etc.).

## 4. Mode de qualification des données et résultats

L'échantillon aléatoire total sur les 3 groupes comprend 390 auteurs-inventeurs candidats parmi les 2501 traités par IPCCAT. Le risque d'homonymie est évalué avec une marge d'erreur < 5% (Thelwall 2009, 29-31). La vérification manuelle sur les sites universitaires et les réseaux sociaux corrobore les effets de la désambiguïsation : 96% sur les 3 corpus. 17 cas font exception (parmi lesquels 15 font doute et moins de 1% relève d'une erreur certaine - soit 0,0005 %)

Le Tableau 1 regroupe ces éléments de qualification :

*Tableau 1 : Résultats de qualification sur les échantillons aléatoires des candidats auteurs-inventeurs*

| Corpus | Univ | Large | LargeBis | Total |
|---|---|---|---|---|
| Inventeurs | 2122 | 1088 | 4510 | 7720 |
| Auteurs-inventeurs candidats (Correspondance IPCCAT seuil > 800) | 1209 | 27 | 1265 | 2501 |
| Echantillon | 137 | 8 | 244 | 390 |
| Candidats vérifiés | 133 (97 %) | 8 | 231 (95 %) | 373 (96 %) |
| Doute | 3 | 0 | 12 | 15 |
| Erreur | 1 | 0 | 1 | 2 (0.0005 %) |

---

[6] L'API IPCCAT est utilisée en général pour identifier le code CIB de textes décrivant des brevets d'invention, ce qu'elle fait avec un très bon niveau de fiabilité d'après les auteurs.
7 Les auteurs d'IPCCAT suggèrent de veiller à un score > 1200 pour garantir la fiabilité des réponses dans le cas des textes de brevets. Notre utilisation détournée d'IPCCAT sur des productions académiques nous conduit à être plus souples quant au seuil utilisé : d'une part le style rédactionnel des articles scientifiques diffère de celui d'un brevet, d'autre part un seuil plus bas permet une tolérance plus vaste quant à la connexité des travaux.

Le Tableau 2 ci-après synthétise les résultats du processus de levée d'homonymie sur les trois corpus. Dans le domaine médical, avec la base PubMed, la correspondance IPCCAT sur **Univ** et **LargeBis** (càd sans imposer la présence d'un déposant académique) montre que **dans les deux cas** plus de 70% des inventeurs homonymes sont candidats auteurs-inventeurs (càd probablement académiques).

*Tableau 2 : Inférence des proportions auteurs-inventeurs par levée d'homonymie et ratio d'autorat*

| Corpus | Univ | Large | LargeBis |
|---|---|---|---|
| Nombre de brevets | 683 | 874 | 3122 |
| Nombre d'inventeurs normalisés uniques | 2122 | 1088 | 4510 |
| Nombre d'auteurs homonymes (avant désambiguïsation) | 1591 | 182 | 1765 |
| Auteurs-inventeurs candidats (Correspondance IPCCAT seuil > 800) | 1209 | 27 | 1265 |
| Proportion d'auteurs-inventeurs matchée (après traitement IPCCAT) sur les auteurs homonymes | 74.78 | 14.89 | 71.67 |

## 5. Conclusion

Ce procédé de désambiguïsation, assimilable à un rapprochement sémantique (Block et al. 2021), établit l'autorat effectif dans plus de 95% des cas, le contrôle manuel sur échantillon aléatoire des auteurs-inventeurs candidats ne révélant que 5% d'erreur. Le groupe **Univ** (déposants académiques) traité avec IPCCAT conserve 74,78% d'homonymes pertinents. Le score similaire (71.67 %) dans le jeu étendu sur les demandes de brevets de l'année précédente (**LargeBis**) montre la stabilité du procédé et sa fiabilité. La focalisation sur une spécialité scientifique est optimisée par le choix d'une base de données source réellement adéquate (ici : PubMed). La méthode expérimentée ici, dans un contexte de généralisation de la science ouverte, pourrait s'étendre à tout autre domaine scientifique lié à l'industrie.

Le rapprochement des demandes de brevet et des publications par ce procédé met en évidence le « prior art », non seulement pour vérifier la nouveauté de ces demandes mais bien plus pour améliorer la visibilité de la composante « recherche » dans l'avènement de technologies nouvelles (Lissoni et Montobbio 2015) sans recourir par défaut aux seules citations bibliographiques (Verbeek et al. 2002 ; Acosta et Coronado 2003 ; Glänzel et Meyer 2003 ; Ahmadpoor et Jones 2017).

Des travaux antérieurs (Reymond, Galliano et Quoniam 2019) nous ont permis d'esquisser la pertinence de la classification internationale des brevets comme pivot de l'interdisciplinarité sur la base des thèses en ligne. Cette deuxième approche, bien plus concluante malgré les limites d'erreurs marginales, laisse en outre envisager des utilisations et extensions en scientométrie en tant qu'instrument fiable de décloisonnement des sciences, de l'industrie et de la société.